\documentstyle[epsf,12pt,rotating]{article}
\begin{document}

\catcode`@=11
\long\def\@caption#1[#2]#3{\par\addcontentsline{\csname
  ext@#1\endcsname}{#1}{\protect\numberline{\csname
  the#1\endcsname}{\ignorespaces #2}}\begingroup
    \small
    \@parboxrestore
    \@makecaption{\csname fnum@#1\endcsname}{\ignorespaces #3}\par
  \endgroup}
\catcode`@=12
\newcommand{\newc}{\newcommand}
\newc{\gsim}{\lower.7ex\hbox{$\;\stackrel{\textstyle>}{\sim}\;$}}
\newc{\lsim}{\lower.7ex\hbox{$\;\stackrel{\textstyle<}{\sim}\;$}}
\newc{\gev}{\,{\rm GeV}}
\newc{\mev}{\,{\rm MeV}}
\newc{\ev}{\,{\rm eV}}
\newc{\kev}{\,{\rm keV}}
\newc{\tev}{\,{\rm TeV}}
\newc{\mz}{m_Z}
\newc{\mpl}{M_{Pl}}
\newc{\chifc}{\chi_{{}_{\!F\!C}}}
\newc\order{{\cal O}}
\newc\CO{\order}
\newc\CL{{\cal L}}
\newc\CY{{\cal Y}}
\newc\CH{{\cal H}}
\newc\CM{{\cal M}}
\newc\CF{{\cal F}}
\newc\CD{{\cal D}}
\newc\CN{{\cal N}}
\newc{\eps}{\epsilon}
\newc{\re}{\mbox{Re}\,}
\newc{\im}{\mbox{Im}\,}
\newc{\invpb}{\,\mbox{pb}^{-1}}
\newc{\invfb}{\,\mbox{fb}^{-1}}
\newc{\yddiag}{{\bf D}}
\newc{\yddiagd}{{\bf D^\dagger}}
\newc{\yudiag}{{\bf U}}
\newc{\yudiagd}{{\bf U^\dagger}}
\newc{\yd}{{\bf Y_D}}
\newc{\ydd}{{\bf Y_D^\dagger}}
\newc{\yu}{{\bf Y_U}}
\newc{\yud}{{\bf Y_U^\dagger}}
\newc{\ckm}{{\bf V}}
\newc{\ckmd}{{\bf V^\dagger}}
\newc{\ckmz}{{\bf V^0}}
\newc{\ckmzd}{{\bf V^{0\dagger}}}
\newc{\X}{{\bf X}}
\newc{\bbbar}{B^0-\bar B^0}
\def\bra#1{\left\langle #1 \right|}
\def\ket#1{\left| #1 \right\rangle}
\newc{\sgn}{\mbox{sgn}\,}
\newc{\m}{{\bf m}}
\newc{\msusy}{M_{\rm SUSY}}
\newc{\munif}{M_{\rm unif}}
%
%
\def\NPB#1#2#3{Nucl. Phys. {\bf B#1} (19#2) #3}
\def\PLB#1#2#3{Phys. Lett. {\bf B#1} (19#2) #3}
\def\PLBold#1#2#3{Phys. Lett. {\bf#1B} (19#2) #3}
\def\PRD#1#2#3{Phys. Rev. {\bf D#1} (19#2) #3}
\def\PRL#1#2#3{Phys. Rev. Lett. {\bf#1} (19#2) #3}
\def\PRT#1#2#3{Phys. Rep. {\bf#1} (19#2) #3}
\def\ARAA#1#2#3{Ann. Rev. Astron. Astrophys. {\bf#1} (19#2) #3}
\def\ARNP#1#2#3{Ann. Rev. Nucl. Part. Sci. {\bf#1} (19#2) #3}
\def\MPL#1#2#3{Mod. Phys. Lett. {\bf #1} (19#2) #3}
\def\ZPC#1#2#3{Zeit. f\"ur Physik {\bf C#1} (19#2) #3}
\def\APJ#1#2#3{Ap. J. {\bf #1} (19#2) #3}
\def\AP#1#2#3{{Ann. Phys. } {\bf #1} (19#2) #3}
\def\RMP#1#2#3{{Rev. Mod. Phys. } {\bf #1} (19#2) #3}
\def\CMP#1#2#3{{Comm. Math. Phys. } {\bf #1} (19#2) #3}
\relax
%
%
%
\def\beq{\begin{equation}}
\def\eeq{\end{equation}}
\def\bea{\begin{eqnarray}}
\def\eea{\end{eqnarray}}
%
%
%
\newc{\ie}{{\it i.e.}}          \newc{\etal}{{\it et al.}}
\newc{\eg}{{\it e.g.}}          \newc{\etc}{{\it etc.}}
\newc{\cf}{{\it c.f.}}
%
%
%
%
\def\slash#1{\rlap{$#1$}/} 
\def\Dsl{\,\raise.15ex\hbox{/}\mkern-13.5mu D} 
\def\delsl{\raise.15ex\hbox{/}\kern-.57em\partial}
\def\Ksl{\hbox{/\kern-.6000em\rm K}}
\def\Asl{\hbox{/\kern-.6500em \rm A}}
\def\Qsl{\hbox{/\kern-.6000em\rm Q}}
\def\gradsl{\hbox{/\kern-.6500em$\nabla$}}
%
%
%
\def\bar#1{\overline{#1}}
\def\vev#1{\left\langle #1 \right\rangle}
%

\begin{titlepage}
\begin{flushright}
UND-HEP-01-K02\\
May 2001
\end{flushright}
\vskip 2cm
\begin{center}
{\large\bf Exponential Quintessence and the End of Acceleration}
\vskip 1cm
{\normalsize\bf
Christopher Kolda and William Lahneman}
\vskip 0.5cm
{\it Department of Physics, 225 Nieuwland Hall, University of Notre
Dame, \\ Notre Dame, IN 46556}

\end{center}
\vskip .5cm

\begin{abstract}
Recent observations indicate that the universe's expansion has been
accelerating of late. But recent theoretical work has highlighted the
difficulty of squaring acceleration with the underlying assumptions of
string theory, disfavoring most models of quintessence.
because they predict eternal acceleration. We show that one of the simplest
and most motivated quintessence models described by an exponential
potential can produce the acceleration needed to explain the data
while also predicting only a finite period of acceleration,
consistent with theoretical paradigms. This model is no more tuned
than the canonical tracking quintessence models.
\end{abstract}

\end{titlepage}

\setcounter{footnote}{0}
\setcounter{page}{1}
\setcounter{section}{0}
\setcounter{subsection}{0}
\setcounter{subsubsection}{0}


\section{Introduction}

Over the last several years, new and better sources of astrophysical
data have allowed us to measure the universe's dynamics to an
unprecedented degree. The concensus emerging from 
this data is that we are living
in an (approximately) flat and expanding universe, but one in which
that expansion is either currently, or has in the recent past been,
accelerating~\cite{SN}. 
Such an expansion is quite unexpected, violating our
Newtonian intuition and requiring some form of ``universal repulsion''
to counteract the decelerating efforts of normal matter and radiation.

Modelling the energy content of the universe as 
a perfect fluid, Einstein's equations (in a Friedmann-Robertson-Walker 
universe) yield
$$a(t)\propto t^{\frac{2}{3(1+w)}}$$
where $a(t)$ is the metric scale factor and $w$ defines the pressure
equation of state of the fluid: $p=w\rho$. In order for 
$\ddot a>0$\footnote{Dots will represent time derivatives, primes
derivatives with respect to fields.}
(\ie, acceleration) one needs $w<-\frac13$. Normal matter and
radiation ($w=0,\frac13$ respectively) only lead to deceleration, so
that we are forced to conclude that some new, and unknown, energy
source has come to dominate the energy density of the universe in the
recent past. We will refer to this new
source as the ``dark energy'' henceforth.)
Recent fits favor a universe with roughly 2/3 of its energy in the
form of dark energy, and 1/3 in the form of matter, most of the 
latter being dark matter~\cite{SN}.

The simplest of all sources for the dark energy is a cosmological
constant, $\Lambda$, since $w_\Lambda=-1$. But such an explanation is
not without difficulties. First, the observed energy density today is
roughly $\rho\simeq 10^{-11}\ev^4$ whilst a cosmological constant
would naturally be expected to have a density $10^{120}$ times as
great; values of the size observed would seem to indicate a large
fine-tuning in the underlying field theory.
Second, a cosmological constant suffers from a severe temporal
fine-tuning. While matter and radiation energy densities fall
as $a^{-3}$ and $a^{-4}$, $\rho_\Lambda$ is constant. Why, then,
should we happen to live at that one peculiar time in the history of
the universe at which $\rho_m\approx\rho_\Lambda$? (The question may be
even worse, for one could argue that the proximity of matter-$\Lambda$
equality to matter-radiation equality implies a near triple
coincidence between matter, radiation and $\Lambda$, an event that one
would not expect to occur now nor at any other time in the life of the
universe~\cite{triple}.) Finally, it has been emphasized recently
that universes that accelerate without end have no well-defined
asymptotic states from which a physical S-matrix can be built; thus
it would seem that endless acceleration is in contradiction with the
axioms of string theory~\cite{asymptotic}. 
A cosmological constant, unfortunately, gives just such an eternal
acceleration. 

A large class of models have arisen which attempt to replace
the cosmological constant with something dynamical, in particular, a scalar
field called quintessence~\cite{quint}. 
For a scalar field $\phi$, the equation of state satisfies
\beq 
w=\frac{\frac12\dot\phi^2-V(\phi)}{\frac12\dot\phi^2+V(\phi)}
\eeq
assuming $\nabla\phi=0$. Thus a scalar field
can be used to generate any $w$ with $-1\leq w\leq1$. But most
importantly, the value of $w$ need not remain constant as the universe
expands, thus allowing for the possibility that the value of
$\phi$ today is dynamically determined in such a way as to
explain the small size of $\rho_\phi$ and why we live so close to the
beginning of the $\phi$-dominated era. 

Quintessence models, however, fail to fully solve the problems
associated with the cosmological constant. For one thing, they generally
contain one free parameter which can be taken as their current energy
density, or alternatively, the time at which matter and quintessence
meet; this is little different than just setting the cosmological
constant by hand~\cite{kl}. More recently, it has been observed that most
models cannot solve the last of our
three problems either. That is, once the quintessence field has come
to dominate the universe, a period of acceleration/inflation
begins which has no obvious ending and thus does not allow for a
causal definition of asymptotic particle/string states.

In this paper, we will revive one of the more interesting, though
discarded, models of quintessence and show that it can generate
a period of acceleration which only lasts for a finite time, and that
this model is no more contrived or tuned than any of the standard
``tracking'' quintessence models in the literature~\cite{quint}.
This is the so-called exponential
model, also called the ``scaling'' model for reasons that will soon be
clear.

\section{Exponential Quintessence}

The potential studied in this paper is one of the simplest and most
motivated of the various quintessence potentials put forward:
\beq
V=\hat V e^{-\lambda\phi/M}
\eeq
where $\lambda$ is an unknown coefficient of $\CO(1)$ and
$M=(8\pi G)^{-1/2}$ is the reduced Planck mass. This potential and its
cosmological behavior has been studied already by a number of
authors~\cite{exp}. The behavior of $\phi$ can be studied
by dividing the energy density of the universe into two components: a
``background'' density, $\rho_b$, of either matter or radiation, and
$\phi$ itself, where $\rho_\phi=\frac12\dot\phi^2+V(\phi)=\rho_{\rm
crit}-\rho_b$ (we will assume $\Omega\equiv\frac{\rho_{\rm
tot}}{\rho_{\rm crit}}=\Omega_b+\Omega_\phi=1$
throughout this work). Assuming
no interactions between $\phi$ and ordinary matter or any
self-interactions (consistent with~\cite{kl}), then the
dynamics of $\phi$ are completely determined by the usual cosmological
equations:
\beq
H^2=\frac{1}{3M^2} (\rho_b+\rho_\phi),
\eeq \beq
\dot\rho_b+3H\rho_b(1+w_b)=0,
\eeq \beq
\ddot\phi+3H\dot\phi+V'(\phi)=0,
\eeq
where $H$ is the Hubble constant and 
$w_b=0$ or $\frac13$ for a background of matter or
radiation. Previous authors had discovered a remarkable attractor
solution for $\rho_\phi$:
\beq
\Omega_\phi=\frac{\rho_\phi}{\rho_{\rm crit}}=\frac{3}{\lambda^2}(1+w_b).
\label{omegaphi}
\eeq
That is, there exists an attractive fixed point trajectory on which
the ratio of $\rho_\phi$ and $\rho_b$ is constant at all times.
Thus the energy density of $\phi$ at the attractor solution is
determined completely by $\lambda$ and is independent of $\hat V$. This
also implies that $w_\phi=w_b$ at all points along the attractor. One
rather remarkable implication of this is that $w_\phi$ spontaneously
changes from $\frac13$ to $0$ at the time of matter-radiation
equality. 

A full stability analysis of this model can be
performed~\cite{exp} and the above solution is indeed found to be
an attractor for a wide range of initial conditions under the
assumption that
$\lambda^2>3(1+w_b)$. For smaller $\lambda^2$, this solution
disappears. (It is clear that such a solution would be inconsistent
with our initial assumption of a flat universe.)

Despite its mathematical charm,
it appears that the attractor solution cannot describe dark energy as
we observe it. It has two fundamental flaws. First, in a universe
composed of only matter, radiation and $\phi$, it seems to be
impossible to generate $w_\phi<0$ along the attractor. Thus
acceleration never occurs. The second problem is that in order for
$\phi$ to dominate $\rho$ today, it must have also dominated $\rho$
at all previous times along the attractor. This is grossly
inconsistent with big-bang nucleosynthesis (BBN) constraints; recent
analyses~\cite{mathews} obtain a limit $\Omega_\phi<0.05$ at the time
of BBN. Attempts to resuscitate exponential quintessence include
altering the potential to include a polynomial prefactor 
\beq 
V\propto\left[A+(\phi-B)^\alpha\right]e^{-\lambda\phi/M}
\eeq
which  acts as a potential barrier to $\phi$~\cite{albrecht}. 
One fixes $A$, $B$ and $\alpha$ such that the scaling evolution of $\phi$
was halted in the recent past, pushing $\dot\phi\to 0$ and thus
$w_\phi\to -1$. This solves both of the above problems but at the
price of introducing free parameters which can again be traded for
the value of $\rho_\phi$ today.

We would like to suggest that the solution to the problems inherent in
the exponential quintessence are not nearly so difficult and can 
be addressed without adding any new pieces to the theory while maintaining
the same level of naturalness as exhibited in tracker quintessence models and
in the model of Ref.~\cite{albrecht}. We begin by dividing 
the parameter space of $\lambda$ into three regions:
\begin{description}
\item[Case I:] $\lambda^2<3$\\ For very small $\lambda$ 
there are no attractor solutions for
$\phi$ either during matter- or radiation-domination, 
and thus $\rho_\phi$ is strongly dependent in initial conditions. 
This may not rule out this region of parameters, but it will not 
give the attractive behavior we will be
demonstrating and so we do not consider it further.
\item[Case II:] $3<\lambda^2<4$\\ Here there is no attractor for $\phi$
during radiation domination, but there is an attractor during matter
domination. Thus the universe can be at the attractor solution currently without having to
live on it at all times in the past. We will consider this case in detail below.
\item[Case III:] $\lambda^2>4$\\ This is the usual case with attractors at all
epochs; this case can also be perfectly consistent with observations
as will be discussed below.
\end{description}


It is instructive to begin our discussion with Case II from
above. Here $3<\lambda^2<4$, so that there is no scaling
solution during radiation domination, but there is one during matter domination.
Thus when radiation dominates the universe, the behavior of $\phi$ is
completely controlled by initial conditions. Generically there are two broad
classes of behavior it could follow. If it has a large initial energy
density with $V\gg\dot\phi^2$, 
it will tend to increase, quickly dominating the universe and
leading to inflation without end; this case is uninteresting to us. 
On the other hand, if it has a smaller initial energy
density or $\dot\phi^2\gg V$, then it will tend to fall off rapidly,
as $a^{-6}$, typical for kinetic energy. Thus at early times $w_\phi=1$. 

What halts this rapid drop in quintessence energy? When
$\frac12\dot\phi^2\approx V$, the potential energy begins to dominate
the behavior of $\phi$. This leads to a stabilization of the energy
density at some constant value, that is, $w_\phi=-1$. Following this
constant trajectory, the quintessence energy density then approaches
the background energy density and surpasses it. If the meeting occurs
during matter domination, the attractor solution then kicks in and the
quintessence begins to behave as ordinary matter, $w_\phi=0$. 

This brief history of the quintessence field can most easily be seen in
Figs.~1 and 2. In Fig.~1 we have plotted the energy densities of
matter, radiation, and quintessence as a function of the scale
parameter.  The axes are log-log and have been left unscaled to
emphasize the point that this behavior is independent of the overall
time and energy scales. Fig.~2 shows the behavior of $w_\phi$ as a
function of $\log(a)$. In both figures, we begin at the left during
radiation domination. The quintessence field has an initial condition
that its kinetic energy is much greater than its potential, the latter
being of order that observed today. The quintessence field redshifts
away its energy as $a^{-6}$ until $\frac12\dot\phi^2\approx V$ and
then it changes slope quickly to that of a cosmological constant. 
$\rho_\phi$ remains constant, passing through matter's curve during
the period of matter domination. Once $\rho_\phi$ surpasses $\rho_m$,
the universe begins accelerating.
\begin{figure}
\centering
\turn{-90}
\epsfysize=4.5in
\epsffile{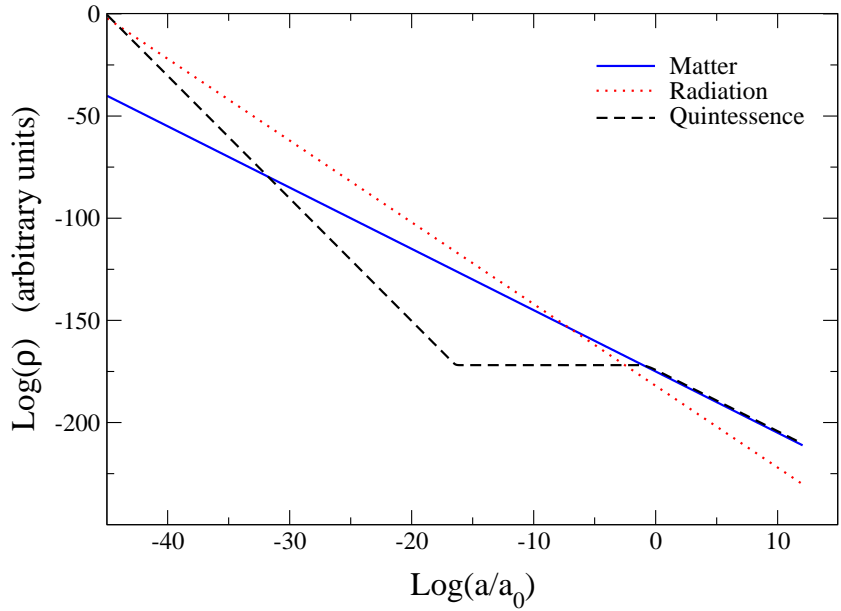}
\endturn
\caption{Scaling of $\rho_r$, $\rho_m$ and $\rho_\phi$ as a function
of scale factor $a$. The current epoch is at $a=a_0$.}
\label{rhofig}
\end{figure}
\begin{figure}
\centering
\turn{-90}
\epsfysize=4.5in
\epsffile{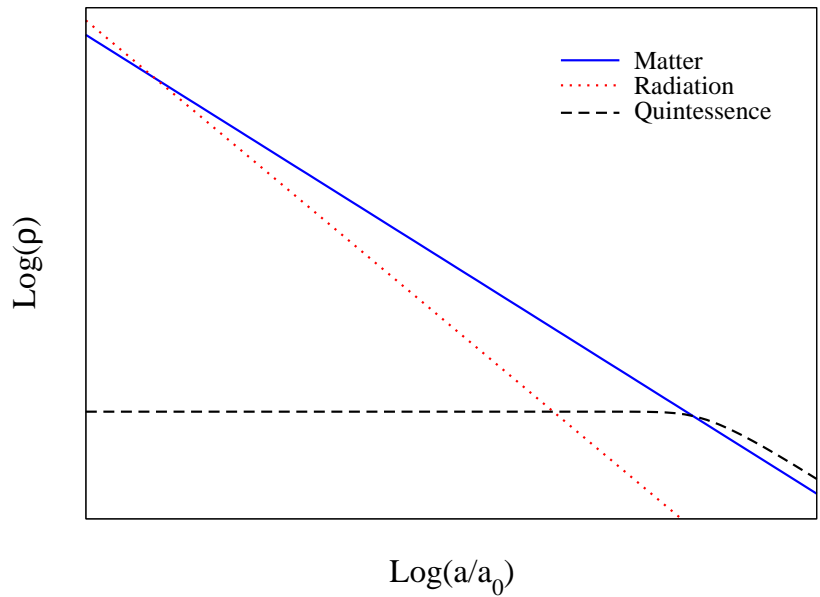}
\endturn
\caption{Detail of Figure~1 showing period of matter domination and 
quintessence domination.}
\label{zoomfig}
\end{figure}
\begin{figure}
\centering
\turn{-90}
\epsfysize=4.5in
\epsffile{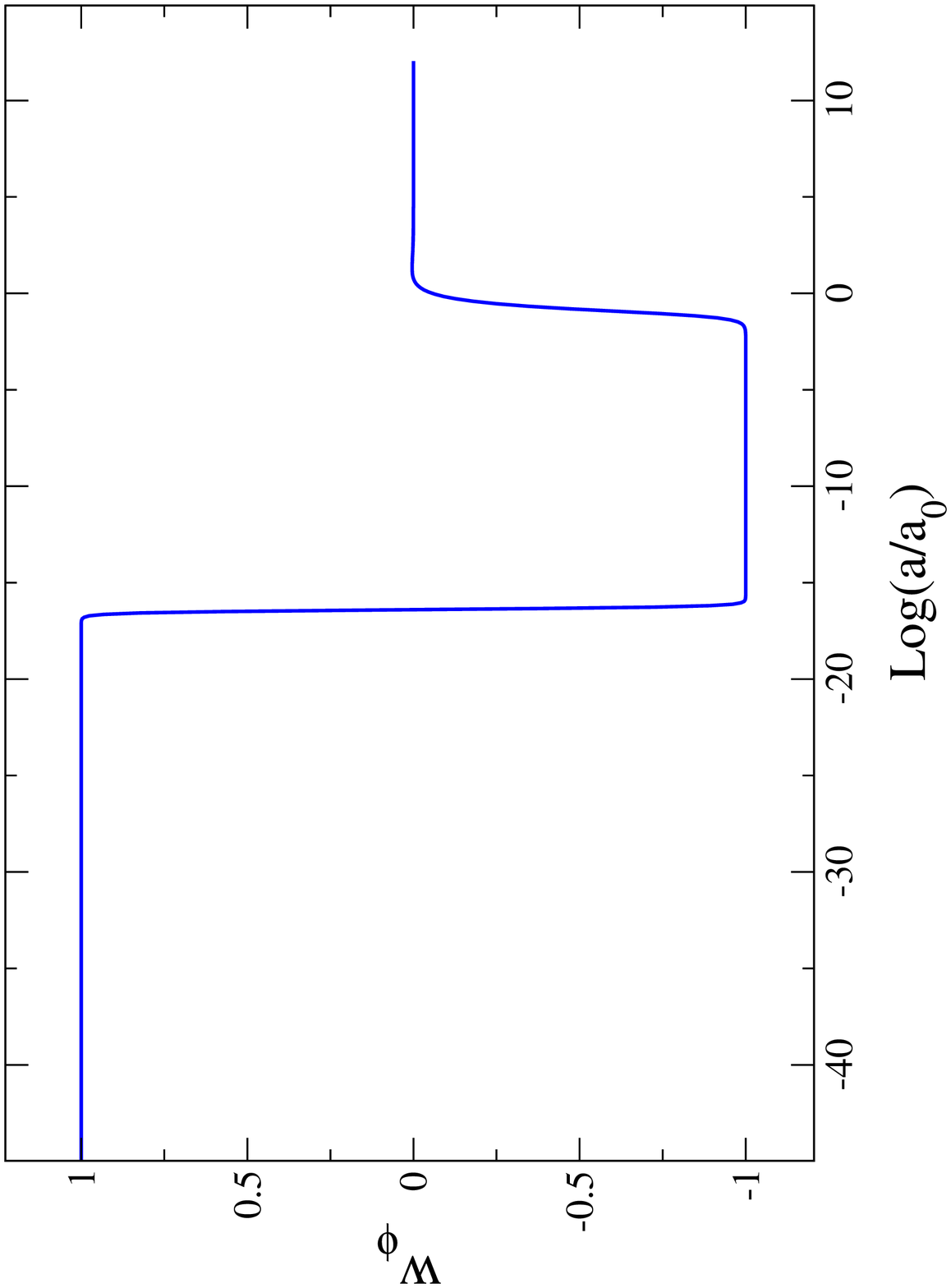}
\endturn
\caption{Equation of state of exponential quintessence from Figure~1.}
\label{wfig}
\end{figure}

However, acceleration only lasts a finite time. Because
$\lambda^2>3$, there exists an attractor solution for $\phi$, in
this case with $\rho_\phi>\rho_m$. Once the attractor is met, $\phi$
turns abruptly to follow it, changing its equation of state to that of
matter, $w=0$, with $\Omega_\phi$ given by Eq.~(\ref{omegaphi}).
And being dominated by a pressure-less field, the
universe discontinues its acceleration. From that point on
the universe acts as though it were matter-dominated, with
the ratio of $\rho_m/\rho_\phi$ fixed perpetually. And string theories
once again have well-defined asymptotic states. 

Notice how this model avoided both of the usual criticisms of
exponential models. First, we were able to generate $w<-\frac13$ {\sl
because we were not yet on the attractor at the time of
matter-quintessence equality!} And again for the same reason, we were
able to avoid any bounds coming from BBN. In fact, it is clear from
the figure that during BBN, the quintessence energy density was many
orders of magnitude too small to affect the expansion rate. The role
of the attractor has changed completely from the usual case in the
literature. Previously one used the attractor to wipe out any dependence
on initial conditions at times long before the present. Here one uses
the attractor to generate an end to the acceleration.

Let us step back for a moment to discuss naturalness. All models of
quintessence of which we are aware 
require (at least) one parameter to be tuned in order to get
matter-quintessence equality at the 
present time. This model will require exactly that same single input;
that is, we will need to input 
the value of the energy density at which matter and quintessence meet:
\beq
V_0=\hat V e^{-\lambda\phi_0/M}
\eeq
Since the values of $\hat V$ and $\phi_0$ can be traded for each
other, some combination of the two must be set to give the observed
dark energy density today. Again, this is no worse than any other
quintessence model. 

Of course, there is now some dependence on initial conditions. For one
thing, we require that at some ``initial'' time, $\dot\phi^2\gg
V$. This is simple to imagine, especially given the small value of $V$
needed to reproduce the current dark energy density. We have solved
this model for a large number of initial values of $\dot\phi$ and find
very little dependence in the final matter-quintessence equality
point. For example, changing $\dot\phi_{\rm initial}$ from $M^2$ to 
$10^{-10}M^2$ resulted in a change of only $10^6$ in the energy
density at matter-quintessence equality.
Thus the dependence on initial conditions is present but not
particularly strong, apart from the
usual need to tune $\hat V$ and/or $\phi_{\rm initial}$ 
to reproduce $V=V_0$ today.

We can learn something of the details of the model by carefully
measuring $\Omega_\phi$ today.
For $3<\lambda^2<4$, the attractor value for
$\Omega_\phi$ falls in the range:
\beq
\frac34<\Omega_\phi<1\quad\Longrightarrow\quad
0<\frac{\rho_m}{\rho_\phi}<\frac13. \quad\quad\mbox{(Case II)}
\eeq
Thus if the observational data settles on $\Omega_{\phi,0}<\frac34$,
we could
conclude that we are still living during the accelerating phase of the
universe's expansion. If however the data settles on
$\Omega_{\phi,0}>\frac34$, it is possible that we have already exited
the accelerating phase. Though we would still be dominated by $\phi$,
the dynamics of the universe as a whole would mimic matter-domination.

How is Case III going to be different? In fact, it is not very
different at all. As long as $V\approx V_0$ at some initial time,
$\phi$ will not reach its attractor solution until the present. This
is true for any $\lambda^2>3$. The only difference between these
cases is the limit on $\Omega_\phi$:
\beq
0<\Omega_\phi<\frac34 \quad\Longrightarrow\quad
\frac13<\frac{\rho_m}{\rho_\phi}<\infty.\quad\quad\mbox{(Case III)}
\eeq
Unfortunately this means that a measurement of $\Omega_\phi<\frac34$
will not by itself tell us whether we have exited acceleration yet.
Observational signals for differentiating this model from others and
for determining the value of $\lambda$ will be the subject of a future
work.

\section{Summary}

The exponential model of quintessence has been often overlooked in the
discussion of viable quintessence models because its attractor
solution does not provide acceleration and is inconsistent with BBN
constraints. However, we have shown here that viable quintessence
models can be built from the exponential potential, but using the
attractor solution as a means for {\sl exiting}\/ 
acceleration, not the converse. One of the key observations of this
work is that the period of acceleration can be generated as the
quintessence field falls onto its fixed point trajectory. However this
behavior is transitory, ending as soon as the fixed point is
reached. This model then can explain acceleration, the large ratio of
dark energy to matter in the present universe, and still remain
consistent with string theory because its acceleration is
short-lived. 

Finally, we explained that this model is no more fine-tuned than other
quintessence models, such as tracking models. There is essentially one
free parameter, which can be taken to be $\hat V$, which sets the
current value of $\rho_\phi$ and thus the time of matter-quintessence
equality. Then for a very wide range of initial conditions on
$\dot\phi$, $\rho_\phi=\,$constant at the time of
matter-quintessence equality, leading to the required acceleration. In
short, this model reproduces all the successes of standard
quintessence while remaining consistent with the underlying
assumptions of string theory.

\section*{Acknowledgements}
The work contained herein was originally
presented as the undergraduate honors thesis of WL at the University of Notre
Dame. While this version of the work was being drafted, a similar work
presenting many of the same arguments concerning exponential quintessence
appeared~\cite{cline}.

\end{document}